\documentclass[12pt]{umj}

\usepackage[english,russian]{babel}
\usepackage{amsmath}%
\usepackage{cite}%
\newtheorem{theorem}{Theorem}

\usepackage{amsfonts, srcltx}

\def\Refname{References}
\firstpage{1}

\usepackage{hyperref}

\def\eq#1{\begin{equation}#1\end{equation}}
\def\eqs#1{\begin{eqnarray}#1\end{eqnarray}}
\def\eqn#1{\begin{equation}\begin{split}#1\end{split}\end{equation}}
\def\seq#1{\begin{equation*}#1\end{equation*}}
\def\seqs#1{\begin{equation*}\begin{split}#1\end{split}\end{equation*}}
\def\qed{\vrule height0.6em width0.3em depth0pt\medskip}

\def\dfrac#1#2{\frac{\partial #1}{\partial #2}}

\def\proof{\noindent {\bf Proof.\ }}
\newcounter{transf}
\newcounter{meq}

\begin{document}
\title[On a series of Darboux integrable discrete equations \dots]
{On a series of Darboux integrable discrete equations on the square lattice}

\author{R.N. Garifullin, R.I. Yamilov}

\address{Garifullin Rustem Nailevish,
\newline\hphantom{iii} Institute of Mathematics, Ufa Scientific Center, RAS,
\newline\hphantom{iii} Chenryshevsky str. 112,
\newline\hphantom{iii} 450008, Ufa, Russia}

\email{rustem@matem.anrb.ru}

\address{Yamilov Ravil Islamovish,
\newline\hphantom{iii} Institute of Mathematics, Ufa Scientific Center, RAS,
\newline\hphantom{iii} Chenryshevsky str. 112,
\newline\hphantom{iii} 450008, Ufa, Russia}

\email{RvlYamilov@matem.anrb.ru}

\thanks{\sc Р.Н. Гарифуллин, Р.И. Ямилов, On a series of Darboux integrable discrete equations on the square lattice}
\thanks{\copyright \ Гарифуллин Р.Н., Ямилов Р.И. 2019}
\thanks{\it Поступила ??? ?? 201? г.}

\maketitle {\small
\begin{quote}
\noindent{\bf Abstract.}
We present a series of Darboux integrable discrete equations on the square lattice. Equations of the series are numbered with natural numbers $M$. All the equations have a first integral of the first order in one of directions of the two-dimensional lattice. The minimal order of a first integral in the other direction is equal to $3M$ for an equation with the number $M$. 

In the cases $M=1,\ 2,\ 3$  we show that those equations are integrable in quadratures. More precisely, we construct their general solutions in terms of the discrete integrals. 

We also construct a modified series of Darboux integrable discrete equations which have in different directions the first integrals of the orders $2$ and $3M-1$, where $M$ is the equation number in series. Both first integrals are unobvious in this case.

\medskip
\noindent {\bf Keywords:} discrete quad-equation, Darboux integrability, general solution

\medskip
\noindent{\bf Mathematics Subject Classification: }{39A14, 39A10, 35L10 }

\end{quote}
 }

\section{ Introduction}
We consider here discrete equations of the form:
\eq{(u_{n+1,m+1}+1)(u_{n,m+1}-1)=\theta(u_{n+1,m}+1)(u_{n,m}-1)\label{ser_th},} where $n,m\in\mathbb{Z}$ and $\theta$ is a constant coefficient. Two integrable equations of this form are known. The case $\theta=1$ is presented in \cite{s10} and the case $\theta=-1$ has been found in \cite{ggy18}. Equations in both the cases are Darboux integrable. Both equations have the first order first integral in the $n$-direction, while in the $m$-direction the minimal orders of first integrals are $3$ and $6$. 

We present a series of equations of the form \eqref{ser_th} with special coefficients $\theta=\theta_M,\ M\in\mathbb{N}$, including two above examples. All the equations are Darboux integrable and have a first integral of the first order in the $n$-direction. The minimal order of a first integral $W_{2,M}$ in the $m$-direction is equal to $3M$ for an equation with the number $M$. 
So, those equations may have in the $m$-direction first integrals of an arbitrarily high minimal order.

A few similar series of integrable equations are known in the literature. In \cite{gy12u} a series of Darboux integrable discrete equations is discussed which, however, are of Burgers type. The minimal orders of first integrals in both directions may be arbitrarily high there. An analogous series of the continuous hyperbolic type equations is discussed in \cite{zs01}. In \cite{gy19} a series of the sine-Gordon type autonomous discrete equations is presented. Autonomous generalized symmetries and conservation laws in both directions may have arbitrarily high minimal orders in this case.  Some series of non-autonomous discrete equations of sine-Gordon type are studied in \cite{ghy15}. 

It is interesting to construct the general solutions for equations of the form \eqref{ser_th} under investigation with such a high minimal order of first integrals. Following \cite{gy15,gsy18,gy17}, we succeed to do it in the cases $M=1,2,3$, where first integrals $W_{2,M}$  have the minimal orders 3,6 and 9. In the case $M=1$ the general solution is constructed in an explicit way and coincides with the known \cite{s10}. In the cases $M=2,3$ we show that those equations are integrable in quadratures. This means that one can construct general solutions in terms of the discrete integrals. 

Using non-point transformations invertible on the solutions of discrete equations \cite{y91,s10}, we construct one more series of Darboux integrable discrete equations. Equations of that series have in different directions first integrals of the minimal orders $2$ and $3M-1$, where $M$ is the equation number in the series. Both first integrals are not obvious in this case.

In Section \ref{integr} we introduce a series of discrete equations, prove the Darboux integrability of those equations, and discuss the minimal orders of first integrals. The general solutions in the cases $M=1,2,3$ are constructed in Section \ref{sec_sol}. A modified series of integrable discrete equations is discussed in Section \ref{mod}.

\selectlanguage{english}
\section{Darboux integrability} \label{integr}
We are going to study the following series of discrete equations:
\eq{(u_{n+1,m+1}+1)(u_{n,m+1}-1)=\theta_M(u_{n+1,m}+1)(u_{n,m}-1),\label{ser_M}}
where $\theta_M$ is the primitive root of unit of the degree $M \in \mathbb{N}$. More precisely, $\theta_1=1$ and for $M>1$ one has:
\eq{\theta_M^M=1,\quad \theta_M^j\neq 1,\ 1\leq j\leq M-1.\label{prim}}
For example, all the primitive roots of the degree $M\leq 4$ read: \eqn{\theta_1=1,\quad \theta_2= -1,\quad \theta_3=-\frac12\pm i\frac{\sqrt3} 2,\quad \theta_4=\pm i.\label{our_b1234}}
 For every $M>2$ we have at least two values of $\theta_M$: 
$\theta_M=\exp(\pm 2\pi i/M)=\cos\frac{2\pi}{M}\pm i\sin\frac{2\pi}{M}.$

An equation of the form 
\eq{\label{dis_eq}F(u_{n+1,m+1},u_{n+1,m},u_{n,m+1},u_{n,m})=0} 
is called Darboux integrable if it has two first integrals $W_1,W_2$, such that
\eq{\label{darb1}(T_n-1)W_2=0,\quad W_2=w_{n,m}^{(2)}(u_{n,m+l},u_{n,m+l-1},\ldots,u_{n,m}),}
\eq{\label{darb2}(T_m-1)W_1=0,\quad W_1=w_{n,m}^{(1)}(u_{n+k,m},u_{n+k-1,m},\ldots,u_{n,m}).}
Here $l,k$ are some positive integers,  and $T_n,T_m$ are operators of the shift in the $n$- and $m$-directions, respectively: $T_n h_{n,m}=h_{n+1,m}$, $T_m h_{n,m}=h_{n,m+1}$.
We suppose that the relations (\ref{darb1},\ref{darb2}) are satisfied identically on the solutions of the corresponding equation \eqref{dis_eq}.  

The functions $W_1$ and $W_2$ will be called the first integrals in the $n$- and $m$-directions, respectively.
We assume here that each of the conditions
\eq{\label{cond_WW}\frac{\partial W_1}{\partial u_{n,m}}\neq0,\qquad\frac{\partial W_1}{\partial u_{n+k,m}}\neq0,\qquad\frac{\partial W_2}{\partial u_{n,m}}\neq0,\qquad\frac{\partial W_2}{\partial u_{n,m+l}}\neq0} is satisfied for at least  some $n,m$. The numbers $k,l$ are called the orders of these first integrals $W_1, W_2$, respectively.

The case $M=1$ is known, see equation (4.6) in \cite{s10}. That equation (4.6) is obtained from \eqref{ser_M} with $\theta_M=1$ by the point transformation $v_{n,m}=\frac{1-u_{n,m}}{1+u_{n,m}}.$ It is shown in \cite{s10} that that equation is Darboux integrable by constructing the first integrals in both directions, and its general solution has been found.
The case $M=2$ is known, see equation (51a) in \cite{ggy18}. The first integrals in both directions have been found there for this equation, see relations (53) in \cite{ggy18}.

For any $M$ equation (\ref{ser_M}) has the following first integral in the $n$-direction:
\eq{    W_{1,M} =(\theta_M)^{-m}(u_{n+1,m}+1)(u_{n,m}-1).\label{W1u}}
This is true, as equation \eqref{ser_M} is equivalent to the relation
\eq{(T_m-\theta_M)[(u_{n+1,m}+1)(u_{n,m}-1)]=0.}
Moreover, formula \eqref{W1u} with $\theta$ instead of $\theta_M$ provides the first integral for equation \eqref{ser_th} for any $\theta$.
As for the $m$-direction, we succeed to find the following formula:
\eq{        W_{2,M} =\frac{(u_{n,m+3M}-u_{n,m+M})(u_{n,m+2M}-u_{n,m})}{(u_{n,m+3M}-u_{n,m+2M})(u_{n,m+M}-u_{n,m})},\label{W2u}
    } which provides first integrals for all the equations (\ref{ser_M}). It is easy to see that these integrals (\ref{W1u}) and (\ref{W2u}) have the orders $1$ and $3M$, respectively. Conditions \eqref{cond_WW} are satisfied for all $n,m$ in this case. The fact that formula \eqref{W2u}  defines a first integral in the case $M=1$ is checked by direct calculation.

\begin{theorem}
The function $W_{2,M}$ defined by \eqref{W2u} is the first integral of equation (\ref{ser_M},\ref{prim}) in the $m$-direction for any $M>1$. 
\end{theorem}

\proof  
At first we denote
\eq{\Psi_{n,m}=(u_{n+1,m}+1)(u_{n,m}-1).}From \eqref{ser_M} we have
\eq{T_m \Psi_{n,m}=\theta_M\Psi_{n,m},\label{eqPsi}} and therefore \eq{T^M_m \Psi_{n,m}=\Psi_{n,m}\label{eqPsi1}} for equations \eqref{ser_M} satisfying condition \eqref{prim}.
By using the notation \eq{u^{(j)}_{n,k}=u_{n,Mk+j},\quad 1\leq j\leq M,\label{zamj}} we rewrite \eqref{eqPsi1} as a system:
\eq{(u^{(j)}_{n+1,k+1}+1)(u^{(j)}_{n,k+1}-1)=(u^{(j)}_{n+1,k}+1)(u^{(j)}_{n,k}-1),\quad 1\leq j\leq M.\label{sys_M}}
We see that all the equations in \eqref{sys_M} are independent,  and each of them coincides with \eqref{ser_M}, in which $M=1$ and $m$ is replaced by $k$. 

For this reason, for any equation of system \eqref{sys_M}, we can use the first integral $W_{2,1}$ of equation \eqref{ser_M}. As a result we get for  those equations:
\eq{W_{2}^{(j)} =\frac{(u^{(j)}_{n,k+3}-u^{(j)}_{n,k+1})(u^{(j)}_{n,k+2}-u^{(j)}_{n,k})}{(u^{(j)}_{n,k+3}-u^{(j)}_{n,k+2})(u^{(j)}_{n,k+1}-u^{(j)}_{n,k})}.\label{W2j}}
Taking into account \eqref{zamj} we are led to:
\eq{W_{2}^{(j)}=\frac{(u_{n,M(k+3)+j}-u_{n,M(k+1)+j})(u_{n,M(k+2)+j}-u_{n,Mk+j})}{(u_{n,M(k+3)+j}-u_{n,M(k+2)+j})(u_{n,M(k+1)+j}-u_{n,Mk+j})}.}
Denoting $m=Mk+j$, we see that the relation $T_n W_{2}^{(j)}= W_{2}^{(j)}$ implies $T_n W_{2,M}=W_{2,M}$ for any $n,m$. As \eqref{sys_M} is equivalent to \eqref{eqPsi1}, the last relation is satisfied on any solution of equation \eqref{eqPsi1} and hence of \eqref{eqPsi}. \qed

The first integral $W_{1,M}$ obviously has the lowest possible order for any $M$. 
As it is known from \cite{s10,ggy18}, in the cases $M=1$ and $M=2$ the integral $W_{2,M}$ also has the lowest possible order in its direction. The same is true for the case $M=3$ as it follows from:

\begin{theorem}\label{W2_3}
Equation (\ref{ser_M},\ref{prim}) with $M=3$ does not have in the $m$-direction any first integral of the order $l<9$. 
\end{theorem}

In order to prove this theorem, we apply a method described in detail in \cite[Section 2.2]{gy12}. That method uses so-called annihilation operators introduced in \cite{h05} and allows one to find the first integrals. In the framework of that method, the proof comes to direct but cumbersome calculation. 

The following hypothesis seems to be true: first integral \eqref{W2u} of equation (\ref{ser_M},\ref{prim})  has the lowest possible order for any $M>1$.

\section{General solutions}\label{sec_sol}
Here we use and improve a method developed in \cite[Section 5.2]{gy15},\cite[Section 4]{gy17},\cite{gsy18}.
We construct the general solutions for equations \eqref{ser_M} with $M=1,2,3.$ In the case $M=1$ a solution will be explicit and will coincide with a solution of \cite{s10} up to the M\"obius transformation $v_{n,m}=\frac{1-u_{n,m}}{1+u_{n,m}}$. Corresponding calculation will be needed, however, for the cases $M=2,3$. In the cases $M=2,3$, the general solutions will be given in terms of discrete integrals in terminology of \cite{gsy18}, i.e. it will be shown that equations \eqref{ser_M} with $M=2,3$ are solved by quadrature. 

Let us consider an ordinary discrete equation \eq{\label{ab}a_{n+1}-a_{n}=A_n,}  where $a_n$ is an unknown function and $A_n$ is given. We will say that $a_n$ is found by the discrete integration, in analogy with the ordinary differential equation $a'(x)=A(x),$ and the solution $a_n$ of equation \eqref{ab} will be called the discrete integral of $A_n$.

The explicit general solution of equation \eqref{dis_eq} will be called a function of the form $u_{n,m}=\Phi_{n,m}[a_n,b_m], $ where $a_n,b_m$ are arbitrary functions of one variable. Here the square brackets mean that the function $\Phi_{n,m}$ depends on a finite number of the shifts $a_{n+j},b_{m+j}.$ Such a solution must identically satisfy equation \eqref{dis_eq} for all values of the functions $a_n,b_m$. For example, the discrete wave equation 
\seq{u_{n+1,m+1}-u_{n+1,m}-u_{n,m+1}+u_{n,m}=0} has the following general solution:
\seq{u_{n,m}=a_n+b_m.}

Equation \eqref{dis_eq} is solved by quadrature if it has a solution of the form: \eq{u_{n,m}=\Phi_{n,m}\left[a_n,b_m,a_n^{(1)},a_n^{(2)},\ldots,a_n^{(j_1)},b_m^{(1)},b_m^{(2)},\ldots,b_m^{(j_2)}\right],\label{dis_in}}
where $a_n,b_m$ are arbitrary functions and the square brackets mean, as above, that the function $\Phi_{n,m}$ depends on a finite number of the shifts of its arguments. The functions $a_n^{(j)}$ are obtained from $a_n$ by a finite number of applications of the shift operator $T_n$, of the functions of many variables, and of the discrete integrations. The functions $b_m^{(j)}$ are obtained from $b_m$ analogously. So, the functions $a_n^{(j)}, b_m^{(j)}$ and therefore solution \eqref{dis_in} are implicit in a sense.

\subsection{Case $M=1$}

Equation \eqref{ser_M} is equivalent to \eq{(u_{n+1,m}+1)(u_{n,m}-1)=\lambda_n,\label{ord_1}} where $\lambda_n$ is an arbitrary function. This is the discrete Riccati equation, and we need to know a particular solution of it in order to linearize and then to solve it. We cannot solve this equation for a given $\lambda_n$ in general case. We use the fact that the function $\lambda_n$ is arbitrary and replace $\lambda_n$ by another arbitrary function $\alpha_n$ which plays the role of particular solution: \eq{\lambda_n=(\alpha_{n+1}+1)(\alpha_n-1).} 
In accordance with the known method of solving the Riccati equation, we use the transformation
\eq{\label{uv1}u_{n,m}=\alpha_n+\frac{\alpha_n-1}{v_{n,m}}} to get a linear equation for $v_{n,m}$:
\eq{\frac{\alpha_{n+1}+1}{\alpha_{n+1}-1}v_{n+1,m}+v_{n,m}+1=0.\label{eqv1}} 

To solve this equation, it is convenient to introduce a new arbitrary function $\beta_n$ instead of $\alpha_n$ as:
\eq{\label{alphabeta}\frac{\alpha_{n+1}+1}{\alpha_{n+1}-1}=-\frac{\beta_{n+2}-\beta_{n+1}}{\beta_{n+1}-\beta_{n}}.} Here we follow \cite{gy19}, see (39), (40). Representing equation \eqref{eqv1} in the form
\eq{(T_n-1)[(\beta_{n}-\beta_{n+1})v_{n,m}+\beta_n]=0,} we find 
\eq{\label{genv1}v_{n,m}=\frac{\beta_n+\omega_m}{\beta_{n+1}-\beta_n},} where $\omega_m$ is another arbitrary function. Finally, using (\ref{uv1},\ref{alphabeta},\ref{genv1}), we find $u_{n,m}$: 
\eq{u_{n,m}=\frac{\beta_{n+1}-2\beta_n+\beta_{n-1}}{\beta_{n+1}-\beta_{n-1}}-2\frac{(\beta_{n+1}-\beta_n)(\beta_n-\beta_{n-1})}{(\beta_{n+1}-\beta_{n-1})(\beta_n+\omega_m)}.\label{gen_1}}
It is easy to check that the function \eqref{gen_1} satisfies equation \eqref{ser_M} with $M=1$ for any values of the arbitrary functions $\beta_n,\omega_m.$ So we have got the explicit general solution of \eqref{ser_M} with $M=1$.

\subsection{Cases $M=2$ and $M=3$}

Equation \eqref{ser_M} is equivalent to \eq{\label{eqM}(u_{n+1,m}+1)(u_{n,m}-1)=\theta_M^m\lambda_n,} where $\lambda_n$ is an  arbitrary function.
It is convenient to go from \eqref{eqM} to an equivalent system by using the transformation \eqref{zamj}:
\eq{\label{sysM}(u^{(j)}_{n+1,k}+1)(u^{(j)}_{n,k}-1)=\theta_M^j\lambda_n,\quad 1\leq j\leq M.}
Let us note that $j$ is a number of the function $u^{(j)}_{n,k}$, while $n$ and $k$ are the discrete variables. Unlike \eqref{eqM}, the right hand side of equations \eqref{sysM} depends on one discrete variable $n$ only, as in the case of \eqref{ord_1}.

By analogy with the previous case $M=1$, we can introduce functions $\alpha_n^{(j)}$, so that:
\eq{\theta_M^j\lambda_n=(\alpha^{(j)}_{n+1}+1)(\alpha^{(j)}_{n}-1),\quad 1\leq j\leq M.\label{alphaj}}
Now we can apply the transformations
\eq{\label{uvj}u^{(j)}_{n,k}=\alpha^{(j)}_n+\frac{\alpha^{(j)}_n-1}{v^{(j)}_{n,k}}} to get the linear equations for $v^{(j)}_{n,k}$:
\eq{\frac{\alpha^{(j)}_{n+1}+1}{\alpha^{(j)}_{n+1}-1}v^{(j)}_{n+1,k}+v^{(j)}_{n,k}+1=0.\label{eqv2}} 
As above, we introduce functions $\beta^{(j)}_{n}$, such that 
\eq{\frac{\alpha^{(j)}_{n+1}+1}{\alpha^{(j)}_{n+1}-1}=-\frac{\beta^{(j)}_{n+2}-\beta^{(j)}_{n+1}}{\beta^{(j)}_{n+1}-\beta^{(j)}_{n}},\label{betaj}}  and we obtain  
\eq{\label{vj}v^{(j)}_{n,k}=\frac{\beta^{(j)}_n+\omega^{(j)}_k}{\beta^{(j)}_{n+1}-\beta^{(j)}_n},} where $\omega^{(j)}_k$ are arbitrary functions on $k$.

We can write down for $u_{n,k}^{(j)}$ formulae analogues to \eqref{gen_1}. The problem is that, instead of one $n$-dependent arbitrary function $\beta_n$ in the case $M=1$, we have now $M$ functions $\beta_n^{(j)}$ with a complex relationship between them defined by \eqref{alphaj} and \eqref{betaj}. We can solve this problem for $M=2$ and $M=3$ in terms of the discrete integrals.

\medskip
\paragraph{\bf Case $M=2$.} A relation between the functions $\alpha_{n}^{(1)}$ and $\alpha_{n}^{(2)}$ is obtained from system \eqref{alphaj} by excluding $\lambda_n$:
\eq{(\alpha^{(1)}_{n+1}+1)(\alpha^{(1)}_{n}-1)=-(\alpha^{(2)}_{n+1}+1)(\alpha^{(2)}_{n}-1)\label{rela}.} If one of these functions is known, then the second one is found from the Riccati equation. If we replace $\alpha_n^{(j)}$ by $\beta_{n}^{(j)}$ by using
\eq{\alpha_n^{(j)}=\frac{\beta_{n+1}^{(j)}-2\beta_{n}^{(j)}+\beta_{n-1}^{(j)}}{\beta_{n+1}^{(j)}-\beta_{n-1}^{(j)}},} then a relation between the  functions $\beta_{n}^{(1)}$ and $\beta_{n}^{(2)}$ becomes even more complex.

In order to solve this problem, we rewrite \eqref{rela} in the form:
\eq{\frac{\alpha^{(1)}_{n}-1}{\alpha^{(2)}_{n}-1}=-\frac{\alpha^{(2)}_{n+1}+1}{\alpha^{(1)}_{n+1}+1}\label{rela_b}.}
Denoting the left hand side by $\gamma_{n+1}$, we derive a system for $\alpha^{(1)}_{n}$ and $\alpha^{(2)}_{n}$:
\eq{\frac{\alpha^{(1)}_{n}-1}{\alpha^{(2)}_{n}-1}=\gamma_{n+1}, \quad \frac{\alpha^{(2)}_{n}+1}{\alpha^{(1)}_{n}+1}=-\gamma_n,} which is solved as follows:
\eq{\alpha^{(1)}_n=-\frac{\gamma_{n+1}\gamma_n+2\gamma_{n+1}-1}{\gamma_{n+1}\gamma_n+1},\quad \alpha^{(2)}_n=\frac{\gamma_{n+1}\gamma_n-2\gamma_{n}-1}{\gamma_{n+1}\gamma_n+1}.\label{alp_2}}

Now we consider $\gamma_n$ as a new arbitrary function, then the functions $\alpha_n^{(1)}$ and $\alpha_n^{(2)}$ are found explicitly by \eqref{alp_2}. The functions $\beta_n^{(1)}$ and $\beta_n^{(2)}$ are found from \eqref{betaj} by two discrete integrations, as relations \eqref{betaj} can be rewritten in the form: $$(T_n-1)\log(\beta^{(j)}_{n+1}-\beta^{(j)}_{n})=\log\frac{1+\alpha^{(j)}_{n+1}}{1-\alpha^{(j)}_{n+1}} .$$ 

Let us use \eqref{uvj} and \eqref{vj} to derive a formula for the solution $u_{n,m}$:
\eq{u_{n,m}=\chi_{m+1}\left(\alpha^{(1)}_n+\frac{(\alpha^{(1)}_n-1)(\beta_{n+1}^{(1)}-\beta_{n}^{(1)})}{\beta_{n}^{(1)}+\omega_m}\right)+\chi_{m}\left(\alpha^{(2)}_n+\frac{(\alpha^{(2)}_n-1)(\beta_{n+1}^{(2)}-\beta_{n}^{(2)})}{\beta_{n}^{(2)}+\omega_m}\right),\label{unm}} where
\eq{\chi_m=\frac{1+(-1)^m}2,\quad \omega_{2k+1}=\omega^{(1)}_k,\quad \omega_{2k+2}=\omega^{(2)}_k.}
In formula \eqref{unm} we have two arbitrary functions $\gamma_n$ and $\omega_m$, and the functions $\alpha_{n}^{(j)}$ and $\beta_{n}^{(j)}$ are defined as explained above. The functions $\alpha_{n}^{(j)}$ are found explicitly, while the functions $\beta_{n}^{(j)}$ are in quadratures.

\medskip
\paragraph{\bf Case $M=3$.} Relations between the functions $\alpha_{n}^{(j)},\ \ j=1,2,3,$ are in this case:
\seqs{(\alpha^{(2)}_{n+1}+1)(\alpha^{(2)}_{n}-1)=\theta_3 (\alpha^{(1)}_{n+1}+1)(\alpha^{(1)}_{n}-1),\\ (\alpha^{(3)}_{n+1}+1)(\alpha^{(3)}_{n}-1)=\theta_3 (\alpha^{(2)}_{n+1}+1)(\alpha^{(2)}_{n}-1).} 
We rewrite these relations to introduce new functions $\gamma^{(1)}_n$ and $\gamma^{(2)}_n$:
\seqs{\frac{\alpha^{(2)}_{n+1}+1}{\alpha^{(1)}_{n+1}+1}=\theta_3 \frac{\alpha^{(1)}_{n}-1}{\alpha^{(2)}_{n}-1}=\gamma^{(1)}_{n+1},\\ \frac{\alpha^{(3)}_{n+1}+1}{\alpha^{(2)}_{n+1}+1}=\theta_3 \frac{\alpha^{(2)}_{n}-1}{\alpha^{(3)}_{n}-1}=\gamma^{(2)}_{n+1}.} Using the shift operator $T_n$, we get two systems for three functions  $\alpha_{n}^{(j)}$. Solutions of these systems read:
\eqs{\alpha^{(1)}_n=\frac{2(\gamma^{(1)}_{n+1}-\theta_3)}{\gamma^{(1)}_{n+1}\gamma^{(1)}_{n}-\theta_3}-1,\quad \alpha^{(2)}_n=\frac{2\theta_3(1-\gamma^{(1)}_{n})}{\gamma^{(1)}_{n+1}\gamma^{(1)}_{n}-\theta_3}+1,\label{alpha_1}\\
\alpha^{(2)}_n=\frac{2(\gamma^{(2)}_{n+1}-\theta_3)}{\gamma^{(2)}_{n+1}\gamma^{(2)}_{n}-\theta_3}-1,\quad \alpha^{(3)}_n=\frac{2\theta_3(1-\gamma^{(2)}_{n})}{\gamma^{(2)}_{n+1}\gamma^{(2)}_{n}-\theta_3}+1.\label{alpha_2}}

We have to agree two different formulae for the function $\alpha^{(2)}_n$. It is convenient to do it for the following function of  $\alpha_{n}^{(2)}$:
\seq{\frac{\alpha^{(2)}_n+1}{\alpha^{(2)}_n-1}=\frac{\gamma^{(1)}_{n}(\theta_3-\gamma^{(1)}_{n+1})}{\theta_3(\gamma^{(1)}_{n}-1)}=\frac{\theta_3-\gamma^{(2)}_{n+1}}{\gamma^{(2)}_{n+1}(\gamma^{(2)}_{n}-1)}.}We rewrite the last equality in the form
\eq{\frac{\gamma^{(1)}_{n}(\gamma^{(2)}_{n}-1)}{\gamma^{(1)}_{n}-1}=\theta_3\frac{\gamma^{(2)}_{n+1}-\theta_3}{\gamma^{(2)}_{n+1}(\gamma^{(1)}_{n+1}-\theta_3)}} and denote the left hand side by $\delta_{n+1}-1.$ Using the shift $T_n$, we get a system for $\gamma^{(1)}_{n}$ and $\gamma^{(2)}_{n}$, which can be expressed as:
\eq{\gamma^{(1)}_{n}=\frac{\delta_{n+1}-1}{\delta_{n+1}-\gamma^{(2)}_{n}},\label{gamm1}}
\eq{\theta_3\delta_n(\gamma^{(2)}_{n})^2-[(\theta_3-1)\delta_{n+1}\delta_n+\delta_{n+1}+\delta_n+\theta_3^2-1]\gamma^{(2)}_{n}+\theta_3^2\delta_{n+1}=0.\label{gamm2}}

Let us now consider $\delta_n$ as a new arbitrary function. All the other $n$-dependent functions are expressed via it. The functions $\gamma^{(1)}_{n},\gamma^{(2)}_{n}$ are given by (\ref{gamm1},\ref{gamm2}), the functions $\alpha_n^{(1)},\alpha_n^{(2)}$ and $\alpha_n^{(3)}$ are found from (\ref{alpha_1},\ref{alpha_2}), and the functions $\beta_n^{(1)},\beta_n^{(2)}$ and $\beta_n^{(3)}$ are found from \eqref{betaj}. Let us note that the functions $\alpha_n^{(j)}$ and $\gamma_n^{(1)}$ are found explicitly, while the functions  $\beta_n^{(j)}$ are found by two discrete integrations, and $\gamma_n^{(2)}$ is defined implicitly by the quadratic equation. Solutions $u_{n,k}^{(1)},u_{n,k}^{(2)}$ and $u_{n,k}^{(3)}$ of the system \eqref{sysM} are given by (\ref{uvj},\ref{vj}). 

These solutions depend on the arbitrary functions $\delta_n$ and $\omega^{(1)}_k,\omega^{(2)}_k,\omega^{(3)}_k$, see \eqref{vj}.
Let us come back to a solution $u_{n,m}$ of equation \eqref{eqM} with $M=3$, equivalent to equation \eqref{ser_M}, which is given by transformation \eqref{zamj}. 
From the viewpoint of this solution we have two arbitrary functions $\delta_n$ and $\omega_m$, where
$$\omega_{3k+1}=\omega^{(1)}_k,\quad \omega_{3k+2}=\omega^{(2)}_k,\quad \omega_{3k+3}=\omega^{(3)}_k.$$

\section{Modified series}\label{mod}
Here we use a transformation theory developed in \cite{y91,s10}.

Let us rewrite equation \eqref{ser_M} in the form: 
\eq{\frac{u_{n,m+1}-1}{u_{n,m}-1}=\theta_M\frac{u_{n+1,m}+1}{u_{n+1,m+1}+1}\label{ser_r}.}
This allows us to introduce a new function $v_{n,m}$, so that:
\eq{v_{n,m}=\theta_M\frac{u_{n,m}+1}{u_{n,m+1}+1},\quad v_{n+1,m}=\frac{u_{n,m+1}-1}{u_{n,m}-1}.\label{vu}}
Resulting relations can be rewritten as:
\eq{u_{n,m}=\frac{v_{n+1,m}v_{n,m}-2v_{n,m}+\theta_M}{v_{n+1,m}v_{n,m}-\theta_M}\quad u_{n,m+1}=-\frac{v_{n+1,m}v_{n,m}-2\theta_Mv_{n+1,m}+\theta_M}{v_{n+1,m}v_{n,m}-\theta_M}.\label{uv}}
Rewriting these formulae at the same point $u_{n,m+1}$, we get an equation for $v_{n,m}$:
\eq{(v_{n+1,m+1}-1)(v_{n,m}-\theta_M)=\theta_M(1-v_{n+1,m}^{-1})(1-\theta_Mv_{n,m+1}^{-1}),\label{ser_v}} where $\theta_M$ is the primitive root of unit.

For all the equations of the form \eqref{ser_M} we have got transformation \eqref{vu} which is invertible on the solutions of  \eqref{ser_M}. Two series of equations \eqref{ser_M} and \eqref{ser_v} are equivalent up to this transformation. The particular case $M=2$ of equation \eqref{ser_v} is presented in \cite[(3.31)]{ggy18} up to $v_{n,m}\to -v_{n,m}$ together with the first integrals.   In the case $M=1$ we can apply the point transformation $v_{n,m}=1+w_{n,m}^{-1}$ and get the following equation: 
\eq{w_{n+1,m+1}w_{n,m}=(w_{n+1,m}+1)(w_{n,m+1}+1).} This is nothing but the discrete Liouville equation found in \cite{h87}. Its first integrals and general solution have been constructed in \cite[(19)]{as99}.
In the general case, by using transformation \eqref{vu}, we can rewrite the first integrals.
 
\begin{theorem}
For any $M\geq 1$ equation \eqref{ser_v} has the following first integrals in the $n$- and $m$-directions, respectively:
\eqs{W_{1,M}&=&\theta_M^{-m}\frac{(v_{n+2,m}-1)v_{n+1,m}(v_{n,m}-\theta_M)}{(v_{n+2,m}v_{n+1,m}-\theta_M)(v_{n+1,m}v_{n,m}-\theta_M)},\label{W1v}\\
W_{2,M}&=&\frac{\left(V_{n,m+2M}^{(M)}V^{(M)}_{n,m+M}-1\right)\left(V^{(M)}_{n,m+M}V^{(M)}_{n,m}-1\right)}{\left(V^{(M)}_{n,m+2M}-1\right)V^{(M)}_{n,m+M}\left(V^{(M)}_{n,m}-1\right)},\label{W2v}\\\nonumber &V^{(M)}_{n,m}&=v_{n,m}v_{n,m+1}\ldots v_{n,m+M-1}.}
\end{theorem}

\proof First integral \eqref{W1v}  is obtained from \eqref{W1u} by direct calculation, using transformation \eqref{uv}.

In order to derive \eqref{W2v}, we need  auxiliary relation. 
It follows from the first relation of \eqref{vu} that for any $k\geq 1$ one has:
\seq{\frac{u_{n,m}+1}{u_{n,m+k}+1}=\frac{u_{n,m}+1}{u_{n,m+1}+1}\frac{u_{n,m+1}+1}{u_{n,m+2}+1}\ldots \frac{u_{n,m+k-1}+1}{u_{n,m+k}+1}=\theta_M^{-k}v_{n,m}v_{n,m+1}\ldots v_{n,m+k-1}.}
Now first integral \eqref{W2u} can be rewritten in the form:
\seqs{W_{2,M}=&\frac{[(u_{n,m+3M}+1)-(u_{n,m+M}+1)][(u_{n,m+2M}+1)-(u_{n,m}+1)]}{[(u_{n,m+3M}+1)-(u_{n,m+2M}+1)][(u_{n,m+M}+1)-(u_{n,m}+1)]}\\=&\frac{\left(\frac{u_{n,m+3M}+1}{u_{n,m+M}+1}-1\right)\left(1-\frac{u_{n,m}+1}{u_{n,m+2M}+1}\right)}{\left(\frac{u_{n,m+3M}+1}{u_{n,m+2M}+1}-1\right)\left(1-\frac{u_{n,m}+1}{u_{n,m+M}+1}\right)}\\=&\frac{\left(\frac{\theta_M^{2M}}{V^{(M)}_{n,m+M}V^{(M)}_{n,m+2M}}-1\right)\left(1-\theta_M^{-2M}V^{(M)}_{n,m}V^{(M)}_{n,m+M}\right)}{\left(\frac{\theta_M^M}{V^{(M)}_{n,m+2M}}-1\right)\left(1-\theta_M^{-M}V^{(M)}_{n,m}\right)}.} As $\theta_M^M=1$, we are led to first integral \eqref{W2v}. \qed

The order of first integral \eqref{W1v} is equal to two. Let us show that this order is minimally possible. If in the $n$-direction there exists a first integral $\widetilde{W}_{1,n,m}(v_{n+1,m},v_{n,m})$ for equation \eqref{ser_v}, then we use transformation \eqref{vu} and get a first integral for equation \eqref{ser_M} of the nonstandard form $\widehat{W}_{1,n,m}(u_{n,m+1},u_{n,m})$ satisfying the relation $(T_m-1)\widehat{W}_{1,n,m}=0$. It is easy to check that this is impossible.

The order of first integral \eqref{W2v} is equal to $3M-1$. In the cases $M=1$ and $M=2$, the fact that this order $3M-1$ is minimally possible follows from \cite{as99} and \cite{ggy18}, respectively. In the case $M=3$ we can prove the same, using the fact that the order $9$ of corresponding first integral \eqref{W2u} of equation \eqref{ser_M} is minimal, see Theorem \ref{W2_3}. 

In fact, in the case $M=3$ let us suppose that equation \eqref{ser_v} has a first integral in the $m$-direction $$\widetilde{W}_{2,n,m}(v_{n,m+k},v_{n,m+k-1},\ldots,v_{n,m})$$ of an order $1\leq k\leq 7$. This means that for some $n,m$
$$\dfrac{\widetilde{W}_{2,n,m}}{v_{n,m}}\neq0,\quad \dfrac{\widetilde{W}_{2,n,m}}{v_{n,m+k}}\neq 0.$$ By using the first relation of \eqref{vu}, we rewrite $\widetilde{W}_{2,n,m}$ in terms $u_{n,m+j}$ and get a first integral for equation \eqref{ser_M} of the following form: $$\widehat{W}_{2,n,m}(u_{n,m+k+1},u_{n,m+k},\ldots,u_{n,m}).$$ It easy to prove that its order equals $k+1$, where $2\leq k+1\leq8<9$, but this is impossible.  

Finally we remark that, using the results of Section \ref{sec_sol} and the first of transformations \eqref{vu}, we can construct the general solutions for equations \eqref{ser_v} with $M=1,2,3$.

\end{document}